\RequirePackage{fix-cm}
\documentclass[twocolumn]{svjour3}          % twocolumn
\smartqed  % flush right qed marks, e.g. at end of proof
\usepackage{graphicx}
\usepackage[numbers,square,comma,sort&compress]{natbib}
\usepackage{hyperref}
\hypersetup{
		colorlinks=true,	      % false: boxed links; true: colored links
	  linkcolor=blue,          % color of internal links
    citecolor=blue,        % color of links to bibliography
    urlcolor=blue,           % color of external links
}
\journalname{Applied Physics B}
\usepackage{color}

\begin{document}
%\sloppy

\title{An all-solid-state laser source at 671 nm for cold atom experiments with lithium%\thanks{Grants or other notes
}


\author{U. Eismann\and F. Gerbier\and C. Canalias \and A. Zukauskas \and G. Tr\'enec\and J.~Vigu\'e\and F.~Chevy\and C. Salomon}


\institute{U. Eismann\and F. Gerbier\and F. Chevy\and C. Salomon \at Laboratoire Kastler Brossel, ENS, UPMC, CNRS UMR 8552, 24 rue Lhomond, 75231 Paris, France\\
%              Tel.: +123-45-678910\\
%              Fax: +123-45-678910\\
%               Tel.:  +33 01 44 32 34 97\\
%               Fax:  +33 01 44 32 34 34\\
              %\email{eismann@ens.fr}\\
              %E-mail: {\textcolor{blue}{\tt eismann@ens.fr}}\\
              E-mail: {\href{mailto:eismann@ens.fr}{eismann@ens.fr}}\\
\and G. Tr\'enec\and J. Vigu\'e \at
%Laboratoire Collisions Agr\'egats Ractivit\'e Institut de Recherche sur les Syst\`emes
%Atomiques et Mol\'eculaires Complexes,
LCAR, Universit\'e de Toulouse,
Universit\'e Paul Sabatier and CNRS UMR 5589, F-31062 Toulouse, France\\
\and C. Canalias \and A. Zukauskas \at
Department of Applied Physics, Royal Institute of Technology, AlbaNova Universitetscentrum, SE-10691 Stockholm, Sweden\\
}
%}
%\emph{Present address:} of F. Author  %  if needed

\date{Received: date / Accepted: date}

\maketitle

\begin{abstract}
We present an all solid-state narrow line\-width laser source
emitting $670\,\mathrm{mW}$ output power at $671\,\mathrm{nm}$ delivered in a diffraction-limited beam.
The \linebreak source is based on a fre\-quency-doubled diode-end-\linebreak pumped
 ring laser operating on the  ${^4F}_{3/2}
\rightarrow {^4I}_{13/2}$ transition in Nd:YVO$_4$. By using periodically-poled
po\-tassium titanyl phosphate (ppKTP) in an external build\-up cavity,
doubling efficiencies of up to 86\% are
obtained. Tunability of the source over $100\,\rm GHz$ is
accomplished. We demonstrate the suitability of this robust
frequency-stabilized light source for laser cooling of lithium
atoms. Finally a simplified design based on intra-cavity doubling
is described and first results are presented.

% Include keywords, PACS and mathematical subject classification numbers as needed.
\end{abstract}

\section{Introduction}
\label{intro}

The lithium  atomic species is of great interest for cold atom
experiments and the study of quantum degenerate gases. As a member
of the alkali group it offers strong coupling to electromagnetic
fields and a simple level structure including cycling transitions,
thus making it suitable for laser cooling. The significant natural
abundance of fermionic ($^6$Li) as well as bosonic ($^7$Li)
isotopes allows exploration of both sorts of quantum statistics.
The interaction parameter at ultracold temperatures, the s-wave
scattering length, is easily tunable for both species by applying a
DC magnetic field in the vicinity of a Feshbach resonance \citep{Chin2010}. The large width of these resonances,
in addition to the light mass, adds up to the favourable properties of lithium for ultracold atom experiments. %makes lithium a very attractive candidate for quantum gases experiments.

To produce large samples of quantum degenerate gases, one needs
large numbers of pre-laser-cooled atoms in a magneto-optical trap
(MOT). This first step is mandatory before proceeding to the
evaporative cooling phase that leads to quantum degeneracy by
reducing the atom number in favor of phase-space density.  To
optimize the MOT capture process, one usually fixes the laser
intensity around a saturation intensity and uses the available
output power to maximize the beam diameter. Thus, more laser power
leads to a better capture efficiency and larger atom numbers.
 Another important requirement is the quality of the spatial mode needed to efficiently
couple the laser light to single mode (SM) optical fibers.

The wavelength of the lithium D-line resonances ($670.8\,\rm nm$ in air) %\hl{ I am reluctant to use vacuum wavelength, as this is contrary to the tradition of wavelength in air iwhere it is transparent; if we do so here, we must correct the wavelengths which appear in our part..}
currently restricts the choice of light sources to two different kinds of lasers: dye lasers and external cavity diode lasers (ECDLs).
Dye lasers typically deliver watt-level output of monochromatic light in a diffraction limited beam \citep{Johnston1982}. The drawbacks of this technology are an important maintenance effort, high intrinsic phase noise, and the requirement of an expensive pump laser. ECDLs are typically limited to $50\,\mathrm{mW}$ output with limited spatial mode quality, hence further amplification by injection-locked slave lasers or tapered amplifiers is needed to run a cold-atom experiment.

It is thus desirable to develop suitable single-fre\-quency lasers with watt-level
output power. Further applications of such sources include atom interferometry experiments \citep{Miffre2006}, pumping of Cr:LiSAF lasers \citep{Payne1994} and lithium isotope separation \citep{Olivares2002}.

Light sources emitting at $671\,\rm nm$ based on frequency-doubling of $1342\,\rm nm$ Nd:YVO$_4$ or Nd:GdVO$_4$ lasers have been realized previously \citep{Agnesi2002,Agnesi2004,Lue2010,Ogilvy2003,Yao2004,Yao2005,Zhang2005,Zheng2002,Zheng2004,Lenhardt2010}, reaching up to $9.5\,\mathrm{W}$ of cw multi-mode
output \citep{Lenhardt2010}. A solid state single-fre\-quency laser source delivering $920\,\rm mW$ at around $657\,\mathrm{nm}$ has been presented in \citep{Sarrouf2007}.

Here we report on the construction and characterization of an
all-solid-state laser source with $670\,\rm mW$ output power in a TEM$_{00}$
mode operating at $671\,\rm nm$. This is made possible by frequency doubling
a home-made  $1.3\,\mathrm{W}$ Nd:YVO$_4$ $1342\,\mathrm{nm}$
single mode ring laser in an external cavity. The advantages of our source are: watt-level output power in  a single longitudinal and transverse mode with excellent beam
quality, narrow linewidth ($<1\,$MHz), and long-term frequency
stabilization onto the lithium resonance lines. Furthermore,
multi-mode diode laser pumping at 808\,nm is inexpensive and only low-maintenance efforts are required for establishing reliable day-to-day operation.

The paper is organized as follows: in Section\,\ref{las} we describe the infrared single-frequency laser design and results. Section\,\ref{shg} focuses on the frequency doubling of the infrared radiation, whereas Section\,\ref{snl} treats the spectroscopy and frequency-locking systems. In Section\,\ref{lc} the red laser emission is characterized in terms of relative intensity noise, linewidth and longterm stability. In Section\,\ref{icd} we describe a second setup in progress using intra-cavity doubling that has the potential to deliver similar output power at $671\,\rm nm$ while offering higher simplicity of the laser source design.

\section{Infrared laser}
\label{las}
\subsection{Laser setup}
\label{Laser setup}
To realize a single-longitudinal-mode (SLM) laser it is favorable to use a design avoiding standing waves and thus the resulting spatial hole-burning in the active medium \citep{Siegman1986}. %This effect partially circumvents mode competition, thus allowing several longitudinal modes to oscillate simultaneously. Two solutions exist: the twisted-mode linear cavity and the unidirectional ring cavity requiring an optical diode. The first is not feasible using a birefringent gain medium like Nd:YVO$_4$, so the ring cavity configuration was chosen.
Thus we have chosen a ring cavity with proper intra-cavity etalons as frequency-selective elements.

\begin{figure*}
\begin{center}
\includegraphics[width=.9\textwidth]{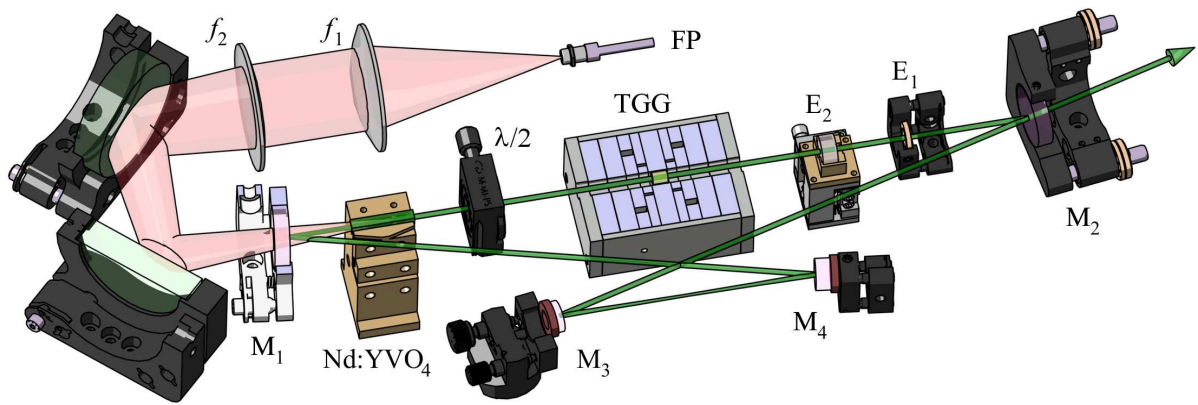}
\caption{(Color online) The laser setup, consisting of the fiber coupled pump source FP, two pump focusing lenses $f_{1,2}$, the cavity mirrors M$_{1-4}$ and the Nd:YVO$_4$ active medium. The $\lambda/2$ wave plate and the TGG Faraday crystal in a magnet ensemble impose unidirectional oscillation whereas two etalons E$_{1,2}$ establish single-mode operation. The optical path of the laser beam is depicted in green. For some mechanical components we show a sectional view to improve visibility of the beam path. The distance between M$_{1}$ and M$_{2}$ is $295\,\rm mm$.
%Some mechanical components are sectioned to improve visibility of the laser beam, depicted in green.}
}
\label{laser}
\end{center}
\end{figure*}

The setup is presented in Fig.\,\ref{laser}. The pump source is a commercial fiber-coupled Coherent FAP-400 diode stack emitting up to $42.6\,\mathrm{W}$ at $808\,\mathrm{nm}$ (90\% energy width: $4\,\mathrm{nm}$). Its metal housing is temperature stabilized to optimize the spectral overlap between pump emission and the gain medium absorption. The fiber output (core radius: $200\,\mathrm{\mu m}$, numerical aperture $\mathit{NA} = 0.22$) is imaged in the gain medium using two lenses ($f_{\rm 1} = 75\,\mathrm{mm}$ and $f_{\rm 2} = 200\,\mathrm{mm}$) to a top-hat spot of radius $w_{\mathrm{pump}} = 530\,\mathrm{\mu m}$. The Nd:YVO$_4$ crystal of dimensions $3\times3\times10\,\rm mm^3$ is 0.2\,at.\%-doped, a-cut and anti-reflective (AR) coated at $808\,\mathrm{nm}$ and $1342\,\mathrm{nm}$. It is wrapped in indium foil and fixed in a solid water-cooled copper mount to efficiently remove heat. Care needs to be taken to avoid acoustic excitations of the mount due to turbulent water flow, thus only a small continuous flux of tap water was applied to prevent frequency fluctuations of the laser output.

The four mirrors M$_{1-4}$ (highly reflective at $1342\,\mathrm{nm}$ except the output coupler M$_2$, transmitting at $808\,\mathrm{nm}$) form a folded ring or bow-tie cavity. The two concave mirrors M$_3$ and M$_4$ have a radius of curvature of $R_{cc} = 100\,\textrm{mm}$.
Thermal design is crucial for the laser: even at moderate pump powers, strong thermal lensing occurs because of the large quantum defect between pump and lasing photon energies and excited state absorption to higher levels \citep{Fornasiero1998,Doualan2008,Okida2005}.  Optimum spatial overlap of the pump fiber image in the laser crystal (top-hat profile of radius $w_{\rm{pump}}$) and laser mode ($1/{\rm e}^2$ radius $w_{\rm{laser}}$) is established by fine-tuning the distance between M$_3$ and M$_4$. The choice of the mode-size ratio $\rho = w_{\rm{laser}} / w_{\rm{pump}} \simeq 1$ allowed for a stable TEM$_{00}$ operation at optimum output power\footnote{Ref. \citep{Chen1997} suggests a ratio $\rho = 0.8$ for which we observe reduced output power as well as higher-order transversal mode oscillation. %However, it is slightly larger than $\rho \simeq 0.8$ suggested by \citep{Chen1997} for which (larger pump spot radius) oscillation of higher-order transverse modes was observed, and output power was reduced.
}. The cavity design remains stable in presence of a thermal focal length down to $f_{\rm th} = 170\,\rm mm$ in the Nd:YVO$_4$ crystal.
Care was taken to design the cavity as short as possible to increase the laser's mechanical stability and the cavity free spectral range (FSR), facilitating SLM behavior.

Unidirectional oscillation is ensured by a combination of a Faraday rotator and a wave plate. The Faraday rotator is custom built by the LCAR group according to the original design presented in \citep{Tr'enec2011}. As a medium displaying the Faraday effect, a cylindrical AR-coated Terbium Gallium Garnet (TGG) crystal is chosen. To minimize absorption, the length of the TGG crystal is limited to ${l_{\mathrm TGG}} = 6\,\rm mm$, and its diameter is $5\,\rm mm$. The ensemble of NdFeB ring magnets delivers a magnetic field integral of $I_B = \int_{0}^{l_{\mathrm TGG}} \! B(z) \, \mathrm{d}z = 8\,\mathrm{T.mm}$ along the TGG axis.
Single-pass measurements resulted in a rotation angle of $\varphi_\mathrm{rot} = -(9.3 \pm 0.1)^\circ$ and thus in a Verdet constant of $\mathcal{V} = \varphi_\mathrm{rot}/I_B = -(20.3 \pm 0.2)\,\mathrm{rad.T^{-1}.m}^{-1}$ for the given crystal at $1342\,\rm nm$. Back rotation and stable unidirectional operation at high intra-cavity powers is established by an AR-coated zero-order $\lambda/2$-wave plate, which is preferred to multi-order wave plates because of instabilities related to thermal effects \citep{Sarrouf2008}. The polarizing intra-cavity element is the Nd:YVO$_4$ crystal which provides higher gain in the c-direction as well as birefringence. The oscillation direction is chosen as indicated in Fig.\,\ref{laser} to spatially separate residual pump light from the output beam.

%\begin{center}
\begin{table}
\caption{Frequency scales in the setup. Typical frequencies are free spectral ranges $\nu_{\rm FSR}$ as described in the text and FWHM for the gain profile from \citep{Doualan2008}. Reflectivities $\mathcal{R}$ are given for the output and input coupling mirrors in case of the two cavities and for the two etalon surfaces, respectively. The finesses $\mathcal{F}$ and Q-factors of the laser cavity and E$_1$ are calculated from the $\mathcal{R}$-values as stated, neglecting further losses. $\mathcal{F}$ and Q were measured for the doubling cavity, see Section\,\ref{dcav}.}
\label{tab:1}       % Give a unique label
\begin{tabular}{lrrc}
\hline\noalign{\smallskip}
 & Typ. freq. & Typ. $\mathcal{R}$ & $\mathcal{F}$/Q-factor  \\
\noalign{\smallskip}\hline\noalign{\smallskip}
Gain profile width & 300\,GHz & & \\
Laser cavity 81\,cm & 360\,MHz & 96.5\% & 110 / $6\times10^7$\\
Doub. cavity 41\,cm & 730\,MHz & 93.6\% & 86 / $3\times10^7$\\
Etalon E$_1$, 0.5\,mm & 210\,GHz & 3.3\% & -/- \\
Etalon E$_2$, 4\,mm & 26\,GHz & 28\% & 2 / 2000\\
\noalign{\smallskip}\hline
\end{tabular}
\end{table}
%\end{center}

Stable SLM behavior could not be established using a single intra-cavity etalon.
Thus, two infrared fused silica etalons E$_{1,2}$ of free spectral range $\nu_{\rm FSR,1} = 210\,\rm GHz$ and $\nu_{\rm FSR,2} = 26\,\rm GHz$ are installed, where $\nu_{\rm FSR} = c/nL_{\rm rt}$, $c$ is the speed of light in vacuum, $n$ the refractive index and $L_{\rm rt}$ the round-trip length.
An overview of the typical frequency scales of the setup is given in Table\,\ref{tab:1}. E$_1$ is non-coated, offering a modulation of the cavity transmission due to its Fresnel reflectivity of $\mathcal{R} = 3.3\%$ per surface, whereas E$_2$ is single-layer coated, yielding  $\mathcal{R} = 28\%$.
%Angular tuning of etalons results in walk-off loss.

A second role of the etalons is coarse frequency tuning of the output radiation. However, angular tuning yields walk-off losses and thus reduces the available output power. By applying the method of \citep{Leeb1975} to ring lasers, the minimum walk-off loss\footnote{The loss estimate of ref. \citep{Leeb1975} yields zero for perpendicular incidence. However, one needs to account for a minimum angle on the order of the Gaussian beam divergence angle to circumvent multi-cavity behavior, causing instability of laser operation.} can be estimated to $\mathcal{L}_{\rm wo} = 0.02\%$ for E$_2$. %($\mathcal{R} = 28\%$).
It can be neglected for E$_1$, for which even an angular tuning of an entire free spectral range only yields $\mathcal{L}_{\rm wo} = 0.03\%$ additional loss. To avoid the higher tilt loss of E$_2$ ($\mathcal{L}_{\rm wo} = 2.8\%$ for angular tuning over one free spectral range of $\nu_{\rm FSR} = 26\,\rm GHz$), we chose to keep it still at the minimum angle and to change its temperature to tune the laser. For that purpose, it is enclosed in a temperature-stabilized copper mount.

Fine tuning of the laser frequency is established by mounting mirrors M$_3$ and M$_4$ on piezoelectric transducers (PZTs): a slow PZT (M$_4$) displaying large displacement of around $2\,\mu$m at maximum voltage of $150\,\rm V$ and a fast PZT (M$_3$) with a displacement of around $\pm 50\,\mathrm{nm}$ limited by the $\pm 15\,\mathrm{V}$ driver.

The laser was mounted on a $50\,\rm mm$-thick breadboard. A combined aluminum-acrylic-glass housing was provided to isolate from acoustic perturbations and for keeping the setup continuously under a dried air atmosphere to prevent dust and moisture from having detrimental effects on stable long-term operation.

\subsection{Laser operation and characteristics}%\label{sec:2}

We now present a detailed description of the laser's operational characteristics.
Pump light absorption in the gain medium depends on the wavelength of the radiation \citep{McDonagh2006}, hence on the pump diode stack temperature. Setting the chiller temperature to $24^\circ\mathrm{C}$ at maximum pump current resulted in highest output power.

\begin{figure}
\includegraphics[width=1\columnwidth]{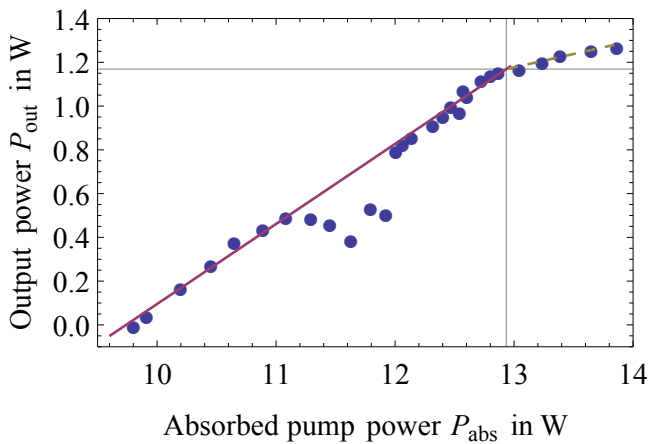}
\caption{(Color online) Single-frequency laser output power as a function of pump current. Two regimes can be distinguished in the data (circles), and linear fits are performed (solid/dashed line) for $P_{abs}<12.9$\,W and $P_{abs}>12.9$\,W. Points in the unstable domain near $P_{\rm abs} = 11.5$\,W were left out of the fit.%, yielding different slopes.}
}\label{pout}
\end{figure}

By choosing a coupling mirror transmission value $\mathcal{T}_{\rm oc} = 3.5\%$, a maximum single-mode output power $P_{\rm{\rm out}}$ of $1.3\,\mathrm{W}$ was obtained, see Fig.\,\ref{pout}.
The lasing threshold was found at an absorbed pump power $P_{\rm{abs}} = 9.8\,\mathrm{W}$. The power rises linearly above threshold with a slope efficiency of $\eta_{\rm{sl}} = \mathrm{d}P_{\mathrm{\rm out}}/\mathrm{d}P_{\mathrm{abs}} = 37\%$. Between $P_{\rm abs} = 11$\,W and 12\,W the output power departs from a linear behavior, and becomes unstable. We attribute this to intracavity-power-induced heating of etalon E$_2$ and the related change of its resonance frequency, thereby leaving the optimum operation range. Above $P_{abs}=12$\,W we recover the initial slope with stable operation.
At $P_{\rm{abs}}(P_{\rm out}) = 12.9\,\mathrm{W}$ $(1.17\,\rm W)$ the slope efficiency drops to $\eta_{\rm{sl}} = 12\%$, indicating the presence of detrimental thermal effects. This behavior was qualitatively found before, see for instance ref.~\citep{Song2002}. Since no degradation of the laser parameters was observed for highest output powers, the laser was always pumped at maximum current for all further measurements.
Removal of the etalons yields a rise in output power of $\sim 20\%$.
This can be partly attributed to detuning of the laser frequency $\omega$ of $\simeq 25\,\mathrm{GHz}$ from the emission peak when lasing at half the lithium resonance frequency.
% at $1342.1\,\rm nm$ and the introduced losses of the additional intracavity elements.

The temperature derivative of the frequency of maximum etalon transmission $\nu_{\rm max}$ yields
\begin{equation}
\frac{\rm{d} \nu_{\rm max}}{{\rm d} T} =  \frac{\nu_{\rm max}}{\nu_{\rm FSR }} \frac{\rm{d}\nu_{\rm{FSR}}} {{\rm{d}}T} = -\nu_{\rm max}\left(\frac{1}{n} \frac{{\rm d}n}{{\rm d}T} + \frac{1}{L_{\rm r}} \frac{{\rm d}L_{\rm r}}{{\rm d}T} \right)\,\mathrm{,}
%\label{}
\end{equation}
where $T$ is the etalon temperature. Putting in the values for IR fused silica from \citep{Leviton2007} yields ${\rm d} \nu_{\rm max}/{\rm d} T = -1.42\,{\rm GHz.K^{-1}}$. The emission wavelength was \linebreak mea\-sured by single-pass frequency doubling the laser light, as described in the next section. This resulted in a second-harmonic (SH) output power in the 1-mW range, sufficient to drive a CCD-based wavelength meter (High Finesse WS-6). The measured temperature dependence of the laser emission frequency $\nu$ is ${\rm d} \nu/{\rm d} T = (-1.45 \pm 0.01)\,{\rm GHz.K^{-1}}$.  Tunability of $\simeq 50\,\rm GHz$ is achieved, yielding $\simeq 100\,\rm GHz$ of tunability for the SH output. To operate the laser at a given frequency without mode hops caused by etalon temperature drifts, the temperature of the etalon $T_{\rm set}$ needs to be stabilized to an interval $T_{\rm set} \pm \delta T$, where $\delta T$ can be estimated to
\begin{equation}
\delta T < \frac{\nu_{\rm FSR, laser}}{2 \left|\textrm{d}\nu/\textrm{d}T\right|} \simeq 0.1^\circ\mathrm{C}\,{\rm .}
\end{equation}
This is accomplished using a homemade temperature controller, offering stability well below this requirement.

Continuous scanning of the laser frequency is \linebreak achieved by sweeping the voltage applied to the slow PZT (M$_4$). For the dynamic range of 0...150\,V this results in more than three times the full mode-hop-free scan range of $\nu_{\rm FSR, laser} = 360\,\mathrm{MHz}$. By applying simultaneous (linear) scanning of the PZT and angle-tuning of etalon E$_2$, mode-hop free continuous frequency tuning of the laser over $1.1\,\mathrm{GHz}$ could be demonstrated, with a resulting maximum output power drop of $\simeq 10\%$ due to etalon walk-off loss.

The beam coupled out of M$_2$ has a $1/\textrm{e}^2$ waist radius of $640\, \rm \mu m$ ($820\,\mu \rm m$) and a divergence angle of $0.8\,\mathrm{mrad}$ ($0.9\,\mathrm{mrad}$) in the horizontal (vertical) plane. The astigmatism results from the cavity design and non-isotropic thermal lensing in the Nd:YVO$_4$ crystal.  %\hl{With the known geometrical cavity parameters this allows to quantify the thermal lens focal length to $f_{\rm th, hor(ver)} = 0.16\,{\rm m}~(0.16\,{\rm m})$ in the horizontal (vertical) direction. The beam quality factor was measured to be $M^2 < 1.1$ in both horizontal and vertical directions using a CCD camera imaging method. This result depends strongly on the cavity geometric parameters which can not be measured with a precision better than 1\,mm, so I think it should be left out of the paper!!}
By employing only spherical lenses, and thus imperfect mode matching, a coupling efficiency to a SM optical fiber of 75\%  was obtained. Stable output power and beam parameters over weeks of daily operation were demonstrated.

\section{Second-harmonic generation}
\label{shg}

\begin{figure*}
\includegraphics[width=1\textwidth]{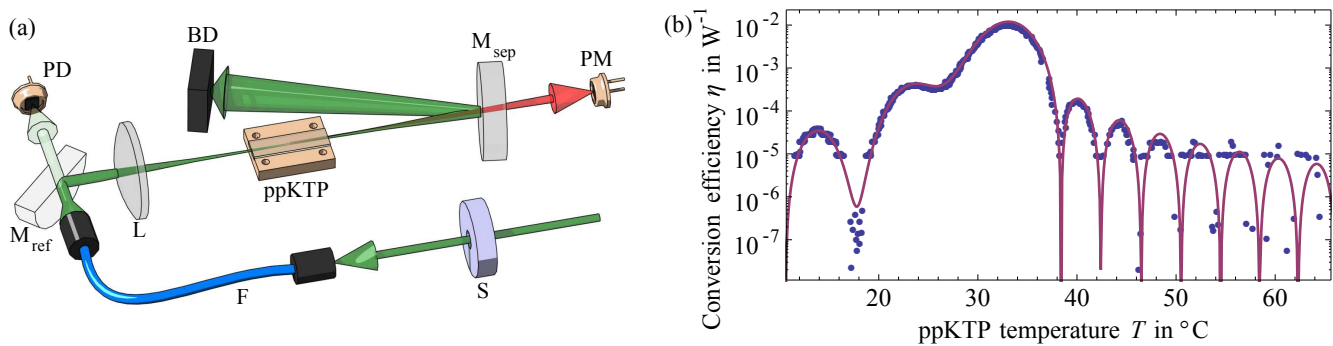}
\caption{(Color online) (a)~Setup of the single pass efficiency measurements. The IR laser output (depicted in green) is mode-cleaned by a polarization-maintaining SM fiber (F). The power leaking through mirror M$_{\rm ref}$ is referenced on a photodiode (PD), and the beam is then focused into the ppKTP crystal by a lens (L). Mirror M$_{\rm sep}$ separates SH (red) from fundamental light, which is sent in a beam dump (BD), whereas the converted light power is measured with a power meter (PM). For referencing to the dark-current values, a shutter periodically switches on and off the IR light. (b)~Temperature-dependent single-pass doubling efficiency. Measured data (circles), fit to eq.\,(\ref{hfunc}) (solid line).}
\label{f:s-pass}
\end{figure*}

\subsection{General considerations}
\label{shgc}

Frequency-doubled light is generated using a second order process in a nonlinear medium. In the limit of weak conversion this yields a second harmonic (SH) output power $P_{2\omega}$ in the form

\begin{equation}
P_{2\omega} = \eta P_\omega^2  \,\rm ,
\label{e:shconv}
\end{equation}
with $P_\omega$ the pump beam power of frequency $\omega$ and $\eta$ the conversion efficiency. In \citep{Boyd1968}, Boyd and Kleinman derived the following expression for $\eta$, assuming Gaussian beams:

\begin{equation}
\eta = \frac{2\omega^3 d_{ij}^2L}{\pi\varepsilon_0 c^4 n_{\omega,i} n_{2\omega,j}}h(\alpha,\beta)\,\mathrm{,}
\label{BK}
\end{equation}
where $d_{ij}$ is the effective nonlinear coefficient of the material with $i(j)$ the polarization of the fundamental (SH) wave, $n_{\omega(2\omega),i(j)}$ the corresponding refractive indices of the material, $L$ the nonlinear material length, $\varepsilon_0$ the vacuum permittivity and $c$ the speed of light in vacuum. The function $h(\alpha,\beta)$ is given as

\begin{equation}
h(\alpha,\beta) = \frac{1}{4\alpha} \left| \int_{-\alpha}^\alpha \! \frac{{\rm e}^{\mathrm{i}\beta(T)\tau}}{1 + \mathrm{i}\tau} \, \mathrm{d}\tau \right|^2\,\mathrm{,}\label{hfunc}
\end{equation}
with  the focusing parameter $\alpha = L/2z_{\rm R}$, where $z_{\rm R}$ is the Gaussian beam Rayleigh length assumed here equal for both waves, yielding a smaller waist for the SH light. The phase-matching parameter

\begin{equation}
\beta = \frac{4\pi z_{\rm R}}{\lambda}\left(n_{\omega,i}(T) - n_{2\omega,j}(T)\right)
\label{ephasem}
\end{equation}
is temperature- and polarization-dependent in the case of birefringent media. The derivation assumes no depletion of the fundamental wave and absence of losses. The integral in eq.\,(\ref{hfunc}) needs to be calculated numerically except for limiting cases, yielding a global maximum of $h_{\mathrm{max}}(2.84, 0.573) = 1.068$. Putting in values for the usual nonlinear media, this results in a doubling efficiency $\eta$ in the \%/W range at best. Thus for the available cw laser power, single pass doubling is not an option and one has to resort to resonantly enhanced intracavity doubling.

Quasi-phase matching in periodically-poled materials is favorable in the intracavity case because of the excellent beam quality achievable without beam walk-off \citep{Myers1995,Boyd2003}. % free process and the avoiding of angle tuning.
The accessibility of the much greater diagonal elements $d_{ii}$ of the nonlinear tensor allows for higher single-pass efficiencies (eq.\,(\ref{BK})) while keeping the phase matching condition of optimum $\beta$.
Compared to the bulk case, the same equations (\ref{e:shconv})-(\ref{ephasem}) hold, by performing the following replacements: $\beta \rightarrow \beta - 2\pi z_{\rm R}/\Lambda$ where $\Lambda$ is the poling period and $d_{ij} \rightarrow d_{\rm{eff}} = 2d_{ii}/\pi$. As a nonlinear medium, periodically-poled Potassium Titanyl Phosphate (ppKTP) was chosen because of its high transparency from 350-4300~nm, its high nonlinear coefficient $d_{33} = 16.9\,\rm pm/V$ \citep{Peltz2001} and its high damage threshold.

\subsection{Single pass measurements}
\label{spm}

We first describe the characterization of the nonlinear crystal using a single-pass method.
The completely automatized measurement setup is represented in Fig.\,\ref{f:s-pass}. The spatial mode-cleaning fiber output power was \linebreak $\simeq 500\,\mathrm{mW}$, resulting in a maximum of  $P_{2\omega} \simeq 2\,\mathrm{mW}$ of red light output, the two beams being separated using mirror M$_{\mathrm{sep}}$.  The fundamental power was monitored using a Ge photodiode (PD) exploiting the finite transmission through mirror M$_{\mathrm{ref}}$. The signal was calibrated against the IR power $P_{\omega}$ hitting the crystal.  The finite transmittance of M$_{\mathrm{ref}}$ at $671\,\mathrm{nm}$ was taken care of, and the SH power measured using a commercial power meter (Thorlabs S130A). The response of the power meter's Si photodiode at $1342\,\mathrm{nm}$ is negligible, and so is the corresponding transmission of M$_{\mathrm{sep}}$. The shutter, driven at $1\,\mathrm{Hz}$ with a 50\% duty cycle allowed for the determination of dark current offset drifts for both power measurements, which is of highest importance for low conversion efficiencies. % and when the free running pump laser modehops, which occurred occasionally.
The crystal is mounted on a transverse ($xy$-) translation stage and temperature controlled to $\sim 10\,\mathrm{mK}$ using a Peltier element and a homemade temperature controller. A set temperature ramp was applied to the controller, scanning the full $55^\circ\mathrm{C}$ range in about 30\,minutes. The slow ramp allowed for adiabatic behavior of the temperature measurement, permitting independent determination of the temperature of the crystal measured by a LM35 sensor attached to the crystal mount.

The ppKTP crystal used in the experiments was fabricated in-house at the Royal Institute of Technology by electric field poling at room temperature \citep{Canalias2006}. Its length is $19.2\,\mathrm{mm}$, featuring an optical aperture of $6\times 1\,\rm{mm}^2$. The length of the per\-iodically-poled region is $17.25\,\mathrm{mm}$. The poling period was chosen to be $\Lambda = 17.61\,\rm \mu m$, resulting in expected plane-wave phase matching at $23.5^\circ\mathrm{C}$ using the tem\-perature-depen\-dent Sellmeier equations from \citep{Fradkin1999,Emanueli2003}. Both surfaces are AR coated at $1342\,\mathrm{nm}$ and $671\,\mathrm{nm}$.

Experimental results are presented in Fig.\,\ref{f:s-pass}. A \linebreak weighted numerical fit to the eq.\,(\ref{BK}) well describes the measured data.  The temperature dependence of the phase matching parameter $\beta$
was taken into account up to quadratic order. The full 99\%-width of the peak of $0.7^\circ\mathrm{C}$ allows the use of standard temperature controllers. However, the optimum phase-matching temperature of $33.2^\circ \mathrm{C}$ differs from the theoretical value. This can be explained by a small difference to the Sellmeier equations as presented in \citep{Emanueli2003} and a non-perfect alignment between pump beam and crystal axis. The maximum measured single pass efficiency of $1.13\%/\mathrm{W}$ represents 74\% of the theoretical maximum of $1.53\%/\mathrm{W}$ from formula (\ref{BK}) with the parameters $\alpha$ as fitted and $d_{33}$ for KTP from \citep{Peltz2001}. This can be explained by imperfections of the domain grating, most probably deviations from the 50\% duty cycle. We thus derive an effective nonlinear coefficient of $d_{\rm eff} = 9.2\,\rm pm/V$ for our crystal.

\subsection{Doubling cavity}
\label{dcav}

\begin{figure}
\includegraphics[width=1\columnwidth]{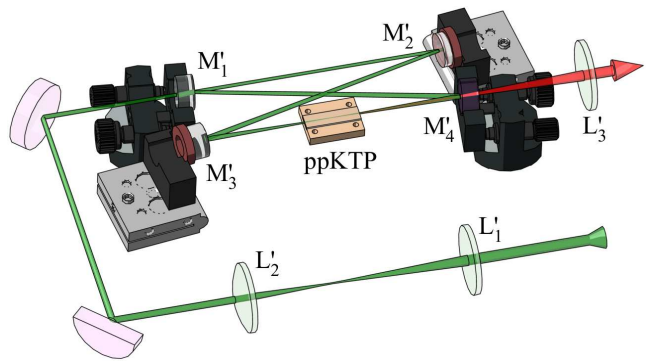}
\caption{(Color online) The doubling cavity setup consisting of the four mirrors M$'_{1-4}$ and the ppKTP nonlinear crystal. The light is coupled to the cavity eigenmode using lenses L$'_{1,2}$, whereas L$'_{3}$ collimates the SH output. Some mechanical components are sectioned to improve visibility of the laser (SH) beam, depicted in green (red). The distance M$'_{3}$--M$'_{4}$ is $95\,\rm mm$, the dimensions of the coupling light paths are not to scale.}
\label{doubling_cavity}
\end{figure}

The doubling cavity setup is similar to the one presented in \citep{Mimoun2008} and depicted in Fig.\,\ref{doubling_cavity}. As for the laser, a four-mirror folded ring-cavity is used, building up a powerful traveling fundamental wave. The pump light is coupled through the plane mirror M$_1'$ for which several reflectivity values $\mathcal{R}_\mathrm{c}$ are available to account for impedance matching. All other mirrors are highly reflective at $1342\,\mathrm{nm}$ and transmitting at $671\,\mathrm{nm}$. M$_3'$ and M$_4'$ are concave with a radius of curvature %R$_{\rm{cc}}$
of $75\,\mathrm{mm}$. M$_2'$ (M$_3'$) was glued on the same type of fast (slow) PZT as used in the laser cavity (Section\,\ref{Laser setup}), allowing to act on the cavity length in the $50\,\mathrm{nm}$\,($2\mu\mathrm{m}$)\,range. The nonlinear crystal is inserted in the cavity's smaller waist of $w_0\simeq55\,\mu\rm m$. The weaker-than-optimal focusing leads to a slightly reduced $h(\alpha = 1.22,0.818) = 0.865$, or a fraction $h/h_{\mathrm{max}} = 81\%$ of the optimum value, yielding $\eta = 0.92\%/\rm W$. This choice represents a trade-off between maximum single-pass doubling efficiency and intensity-related detrimental effects such as nonlinear and SH-induced absorption \citep{Maslov1997} and gray-tracking \citep{Boulanger1999}. % and photorefraction \citep{}.
The cavity length was minimized and the geometry chosen to be shifted with respect to the stability range center. It also avoids frequency degeneracy of higher-order transverse cavity eigenmodes and the TEM$_{00}$ mode. It also accounts for a circular beam in the crystal and thus for a circular SH output. Mode matching between the laser output and the cavity was accomplished using a set of spherical lenses. The crystal mount is identical to the one described in Sect. \ref{spm}. The frequency doubled light is transmitted  through M$_4'$ and collimated using a $f'_3 = 150$~mm lens to a $1/{\rm e}^2$ beam radius of $0.9\,\mathrm{mm}$. The doubling cavity is kept in a housing equaling the laser housing in design and function. % under dry air atmosphere in an aluminum-acrylic-glass housing to prevent dust and moisture from deteriorating the quality of the optical surfaces, and to shield from thermal and acoustic perturbations.

For low intra-cavity powers of up to $\sim 500\,\mathrm{mW}$ and the crystal temperature tuned far from the optimum phase matching value, nonlinear conversion can be neglected. While scanning the cavity length by $\Delta L$, one measures a power signal leaking through M$_2'$, proportional to the intra-cavity power
\begin{equation}
P(\delta L) =  \sum_{lm}{\frac{P_{lm}}{1+F\sin^2{(\varphi_{lm} + \omega\Delta L/c)}}} \,\mathrm{,}
\end{equation}
where $F$ is a fit parameter. $P_{lm}$ are the contributions from the TEM$_{lm}$ modes, displaying a constant cavity round-trip phase of $\varphi_{lm}$, and $c$ is the speed of light in vacuum. The mode matching efficiency is defined as $\eta_{\rm{mo}} = P_{\rm{00}}/\sum{P_{lm}}$. It was maximized to $\eta_{\rm{mo}} = 92\%$. The linear cavity round-trip losses $\mathcal{L}_{\rm emp/tot}$ can be quantified from the fit parameter $F$, where emp(tot) means the empty cavity (cavity including ppKTP crystal). The results are presented in Table\,\ref{t:loss}. Inserting the crystal rises the losses by $\mathcal{L}_{\rm c}\simeq 1\%$. This can be accounted for by residual absorption and scattering in the ppKTP crystal and imperfections of its AR coatings. Taking into account nonlinear conversion, the fundamental mode intracavity power $P_{00} = P_\omega$ at TEM$_{00}$ resonance (referred to as cavity resonance) is a solution of

\begin{equation}
P_\omega = \frac{(1 - \mathcal{R}_{\rm c} - \mathcal{L}_{\rm 1})\eta_{\rm c} P_{\rm p}}{\left( 1 - \sqrt{\mathcal{R}_{\rm c}(1 - \mathcal{L}_{\rm pa}) (1 - \eta P_\omega)}\right)^2}\,{\rm ,}
\label{P00}
\end{equation}
which can be calculated numerically, where $P_{\rm p}$ is the fundamental pump power, $\mathcal{L}_1$ is the coupling mirror (M$_1'$) transmission loss and $\mathcal{L}_{\rm pa}$ is the total cavity passive loss excluding the coupler transmission. The single-pass doubling efficiency $\eta$ is calculated according to eq.\,(\ref{BK}) with $d_{33}$ as measured in Section\,\ref{spm}. % and $h$ as given above.
Setting $\mathcal{L}_{\rm pa} = \mathcal{L}_{\rm c}$ and $\mathcal{L}_{\rm 1} = 0$ the solution of eq.\,(\ref{P00}) yields a maximum SH power of $710\,{\rm mW}$ at the maximum available pump power of $P_{\rm p} = 860\,{\rm mW}$. This is accomplished for an optimized coupling mirror reflectivity of $\mathcal{R}_{\rm c} = 92\%$, yielding a power conversion efficiency of $\eta_{\rm conv} = P_{2\omega}/P_{\rm p} = 84\%$.

\begin{table}
\caption{Passive losses in the doubling cavity measured from cavity transmission spectra at low power and conversion efficiency. $(1 - \mathcal{R}_\mathrm{c})$ is the specified coupler power transmission, $\mathcal{L}_{ \rm emp}$ are the measured empty-cavity round trip power losses, $\mathcal{L}_\mathrm{tot}$ are losses including the ppKTP crystal, and $\mathcal{L}_\mathrm{c}$ are the inferred crystal insertion losses according to  $\mathcal{L}_\mathrm{c}$ = 1-(1-$\mathcal{L}_\mathrm{tot}$)/(1-$\mathcal{L}_{\rm emp}$).}
\label{t:loss}
\begin{tabular}{rrrr}
\hline\noalign{\smallskip}
$1 - \mathcal{R}_\mathrm{c}$ & $\mathcal{L}_{\rm emp}$ & $\mathcal{L}_{\rm tot}$ & $\mathcal{L}_{\rm c}$\\
\noalign{\smallskip}\hline\noalign{\smallskip}
5\% & $6.4\%$ & $7.1\%$ & $0.7\%$ \\
10\% & $10.4\%$ & $11.2\%$ & $0.9\%$\\
17\% & $17.9\%$ & $19.0\%$ & $1.3\%$ \\
\noalign{\smallskip}\hline
\end{tabular}
\end{table}

\begin{figure}
\includegraphics[width=1\columnwidth]{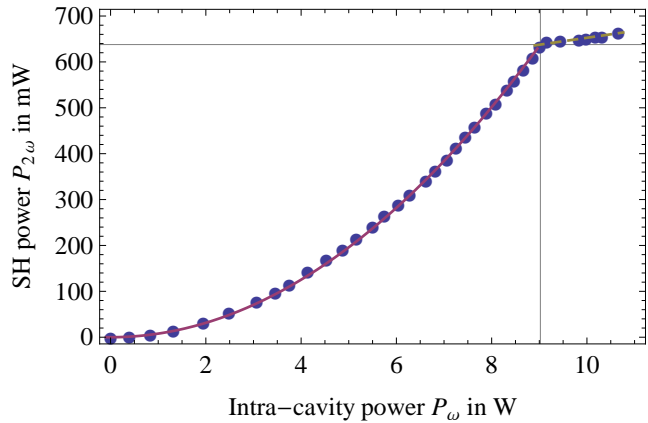}
\caption{(Color online) Intra-cavity conversion for the cavity locked to the laser, as described in Section\,\ref{cavlock}. Measured data (circles) versus parabolic fit (solid line). At $P_{2\omega} = 640\,\mathrm{mW}$ the conversion becomes less efficient (dashed line).}
\label{ic_vs_red}
\end{figure}

After locking the cavity to the laser, as will be described in Section\,\ref{cavlock}, the SH power $P_{2\omega}$ versus $P_{\omega}$ was measured for $\mathcal{R}_\mathrm{c} = 95\%$, see Fig.\,\ref{ic_vs_red}. For low powers, the conversion shows quadratic behavior as stated in eq.\,(\ref{e:shconv}). %Section\,\ref{shgc}.
A fit yields a single-pass efficiency of $\eta = 0.78\%/\rm W$, slightly lower than predicted.
However, starting from the threshold value $P_{\omega}\,(P_{2\omega}) = 9.0\,{\rm W}$ \linebreak $(640\,\mathrm{mW})$, only a slow linear rise in SH power with intra-cavity power is obtained, reaching its maximum at $P_{\omega}(P_{2\omega}) = 10.7\,{\rm W}\,(670\,\mathrm{mW})$. We attribute this behavior to fast intensity-dependent detrimental effects. % mentioned above. %, as indicated
This is confirmed by the lock error signal, which becomes very noisy above threshold. In contrast to \citep{Torabi-Goudarzi2003,Arie1998} the cavity remains locked for all power levels.
When changing the pump power, the SH output follows without observable hysteresis. Long-term degradation is not observed, indicating the absence of gray-tracking. However, for further characterization the setup is operated just below threshold to avoid the related rise in intensity noise. %, displaying optimal intensity noise characteristics.
A maximum doubling efficiency of $\eta_{\rm conv} = 86\%$ is obtained just below threshold, %at $P_{2\omega} = 650\,\mathrm{mW}$,
compatible with the theoretical predictions.

\section{Lock and saturation spectroscopy}
\label{snl}

\begin{figure}
\includegraphics[width=\columnwidth]{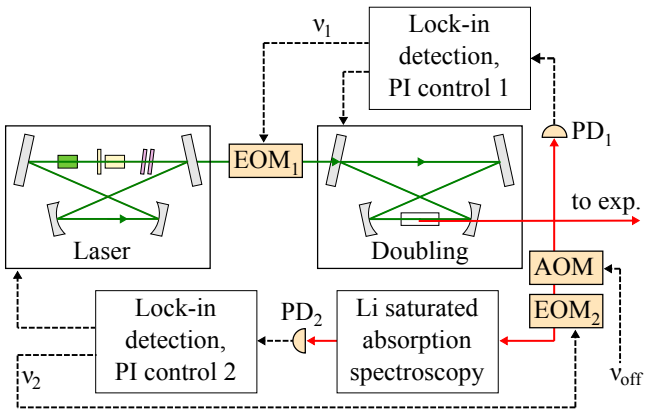}
\caption{(Color online) Locking scheme. Straight lines depict light paths, dashed lines electronic signals. First the doubling cavity is locked to the free-running laser using control circuit components indexed~1. A part of the SH light is used for lithium spectroscopy.
In a second step this reference serves to lock the laser frequency with respect to half of one of the lithium resonance frequencies with a tunable offset, using control circuit\,2. The frequency offset is determined by the double pass AOM driving frequency $\nu_{\rm off}$.}
\label{Lock_scheme}
\end{figure}

Frequency locking of the laser system to the lithium D-line transitions requires frequency doubled light to perform spectroscopy on atomic lithium vapor. Thus, first the doubling cavity needs to be frequency-locked to the free-running laser. In a second step, the laser is stabilized to half the required frequency using lithium saturated absorption phase modulation spectroscopy. The setup is presented in Fig.\,\ref{Lock_scheme}.

\subsection{Cavity lock}
\label{cavlock}

To frequency-lock the doubling cavity to the laser frequency, an error signal needs to be generated. %Two options are common: the H\"ansch-Couillaud \citep{Hansch1980} method or modulation techniques. We use the latter, avoiding zero-offset drifts due to the involved polarizing components. 
We use a modulation technique: An electro-optical modulator (EOM$_1$ in Fig.\,\ref{Lock_scheme}) phase-modulates the infrared pump light at a modulation frequency of $\nu_1 = 1\,\mathrm{MHz}$. In the doubling process, this results in a phase modulation of the frequency-doubled light, which is detected and demodulated using a homemade synchronous detection circuit. It allows to produce an error signal with a 3\,dB-bandwidth of $100\,\mathrm{kHz}$, which is fed into a lock circuit.

The lock circuit combines a proportional-in\-te\-gra\-ting (PI) stage and splits the resulting lock signal in two frequency ranges: $0\,\mathrm{Hz}$ to $\nu_{3\,\rm{dB, slow}} = 72\,\mathrm{Hz}$ for the slow PZT and $72\,\mathrm{Hz}$ to $\nu_{3\,\rm{dB, fast}} = 34\,\mathrm{kHz}$. The amplitude of the slow PZT signal is further amplified by a commercial high-voltage amplifier (Falco Systems WMA-280). The upper frequency limits were chosen to avoid oscillation of the loop at resonances attributable to the PZTs. When scanning the laser frequency via the slow PZT (M$_4$), the ramp signal (modified by an adjustable gain) is fed-forward to the lock signal, thus minimizing lock deviations and stabilizing output power. The implementation of the re-locking scheme of ref. \citep{Mimoun2010} renders the doubling cavity lock significantly more stable to external disturbances.

\subsection{Saturation spectroscopy and laser lock}\label{sll}

A small fraction of the frequency-doubled light is sent through a $200\,\mathrm{MHz}$ acousto-optic modulator (AOM) double-pass setup to frequency-shift the light used for spectroscopy by $2\nu_{\rm mod}$. It is employed to perform saturated absorption spectroscopy in an atomic lithium vapor cell. The required vapor pressure is obtained by heating a metallic lithium sample of natural isotope composition (8\% $^6$Li, 92\% $^7$Li) up to $330^\circ\mathrm{C}$ under vacuum. We use a $50\,\rm cm$ long CF-40 tube with broadband AR-coated windows. The final sections of the tube are water-cooled to prevent from too high temperatures at the CF-40 flanges and windows. Nickel gaskets are employed because of their chemical inertness to lithium vapor. A small amount of argon buffer gas is used to %slow down hot lithium atoms and
force lithium atoms by collision to stick to the side walls before arriving at the window surfaces. The argon pressure is kept low enough to not cause significant collisional broadening of the saturated spectroscopy lines. A metallic mesh put inside the tube covers the tube walls to regain condensed lithium from the colder parts exploiting the temperature-dependent surface tension.

\begin{figure}
\includegraphics[width=1\columnwidth]{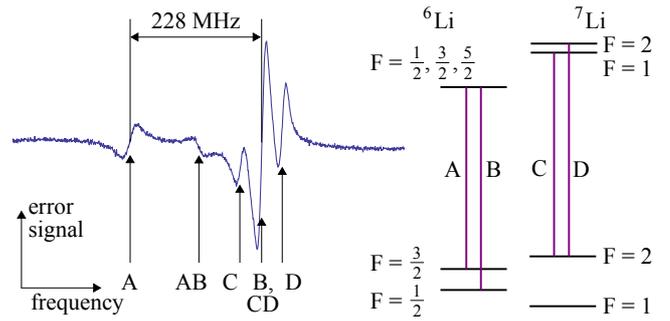}
\caption{(Color online) (a) Lock-in saturated absorption spectroscopy of lithium vapor, and corresponding transitions (b). The transitions are  $2\,^2S_{1/2}\rightarrow 2\,^2P_{3/2}$ for $^6$Li (D2) and $2\,^2S_{1/2}\rightarrow 2\,^2P_{1/2}$ for $^7$Li (D1). Not all levels are shown, and the hyperfine structure of the $^6$Li excited state remains unresolved. Double indexes mark crossover lines.}
\label{Saturated_Absorption}
\end{figure}

The spectroscopy beam $1/\mathrm{e}^2$-radius is $\simeq1\,\mathrm{mm}$, the pump power is of the order of $10\,\mathrm{mW}$, of which typically 50\% are transmitted through the lithium cell on atomic resonance. The beam then passes through a ND filter and an EOM$_2$, which serves to phase-modulate the light at $\nu_2 = 20\,\mathrm{MHz}$. A quarter-wave plate and a mirror retro-reflect the beam with a polarization rotated by $90^\circ$, thus creating the probe beam of $\sim 200\,\mu$W power. Around $100\,\mu$W of probe light are detected on a fast photodiode (Newport 1801). Lock-in detection using a commercial amplifier (Toptica PD110) allows to generate a dispersive error signal. A typical example of a $\simeq 600\,\mathrm{MHz}$-scan over a part of the lithium lines is shown in Fig.\,\ref{Saturated_Absorption}. The hyperfine structure of both lithium isotopes is clearly resolved and error signals of ${\rm SNR}\geq100$ in a 1-MHz bandwidth are detected.  The saturated spectroscopy transmission signal can serve as the auto re-lock reference. This requires a  well pronounced peak or dip structure as satisfied for some of the lines.

To lock the laser frequency with respect to one of the resonances, a two-way PI circuit similar to the one used for locking of the doubling cavity is employed. The AOM frequency and thus the lock offset frequency can be changed by a few $\rm MHz$ while the laser remains locked.

\section{Laser characterization}
\label{lc}

We now present further characterizations of the light source in terms of intensity noise and linewidth. The excellent beam quality of the SH light is confirmed by a SM fiber coupling efficiency of 83\%.

\subsection{Relative intensity noise}

The relative intensity noise spectral density $S_{\rm RIN}(\nu)$ of the SH output was measured by shining a beam of $\sim 120\,\mu W$ on a low-noise photodiode (Newport 1801, \linebreak 125-MHz bandwidth) and recording the signal using a digital oscilloscope (Pico Technology PicoScope 4424) in AC mode, yielding the relative power fluctuations $\varepsilon(t)$ after normalization, where $I(t)/\langle I \rangle_T = 1 + \varepsilon(t)$ with $I(t)$ the intensity and $\langle I \rangle_T$ its temporal average. The definition of $S_{\rm RIN}(\nu)$ is

\begin{equation}
S_{\rm RIN}(\nu) = \lim\limits_{t_{\rm m} \rightarrow \infty}\,{\frac{1}{t_{\rm m}} \left\langle\left| \int_{0}^{t_{\rm m}} \! \varepsilon(t) \mathrm{e}^{{\rm i} 2\pi\nu t} \, \mathrm{d}t \right|^2 \right\rangle }\,\mathrm{}
\end{equation}
with the measurement time $t_{\rm m}$ and $\langle ... \rangle$ denoting temporal averaging. It was realized employing a time-dis\-crete Fourier transformation method and averaging over 100 spectra.

\begin{figure}
\includegraphics[width=1\columnwidth]{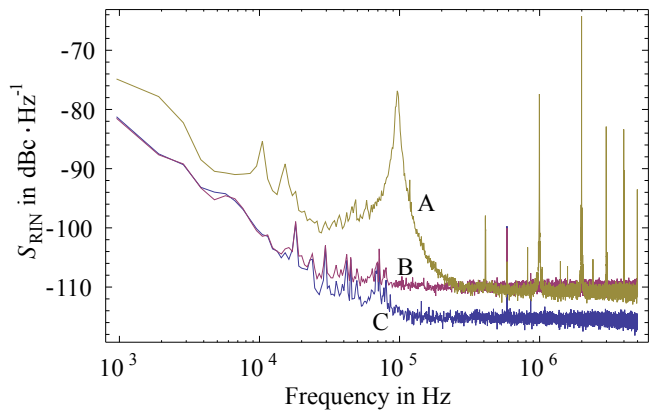}
\caption{(Color online) The SH relative intensity noise spectrum (A), noise for an equivalent photocurrent from a non-coherent source (B) and noise of the detection circuit with no photocurrent (C).}
\label{noise}
\end{figure}

The result is shown in Fig.\,\ref{noise}. The broad peak at $\simeq 100\,{\rm kHz}$ can be attributed to the laser relaxation oscillations. The structure in the $10\,{\rm kHz}$ region can be attributed to the locking system. Above $300\,\rm kHz$ $S_{\rm RIN}$ drops to the photon shot-noise level, as indicated by the spectrum of a noncoherent source producing an equivalent photocurrent (spectrum B in Fig.\,\ref{noise}). The narrow peaks at $1\,\rm MHz$ and harmonics stem from the phase modulation of the pump light, see Section\,\ref{cavlock}. The square root of the integral of $S_{\rm RIN}(\nu)$ from $1\,{\rm kHz}$ to $5\,{\rm MHz}$ ($1\,{\rm kHz}$ to $0.9\,{\rm MHz}$) yields a RMS relative intensity noise of $1.1\times10^{-3}$ ($0.8\times10^{-3}$).

\subsection{Absorption spectroscopy of ultracold atoms}

The laser setup was used as an absorption imaging light source for our lithium quantum gas experiment described elsewhere \citep{Nascimb`ene2009}. A sample of around $1.2\times10^5$ $^7$Li atoms above Bose-Einstein condensation threshold was prepared in an elongated optical dipole trap. Putting a $700\,{\rm G}$ magnetic offset field, the internal electronic states of the atoms are to be described in the Paschen-Back regime. The corresponding lift of degeneracy for the $F=2\rightarrow F'=3$ transition frequencies results in a cycling transition, rendering this method insensitive to constant homogeneous stray fields.
By applying a laser frequency detuning $\delta$ with respect to atomic resonance using the offset lock as described in section \ref{sll}, one detects a different atom number $N(\delta)$ while assuming constant trap conditions according to
\begin{equation}
\frac{N(\delta)}{N(0)} =  \left[1+\left(\frac{2\delta}{\Gamma}\right)^2\right]^{-1}\,\mathrm{,}
\label{natoms}
\end{equation}
where $\Gamma$ is the measured linewidth of the transition and $N(0)$ the atom number detected at resonance.
\begin{figure}
\includegraphics[width=1\columnwidth]{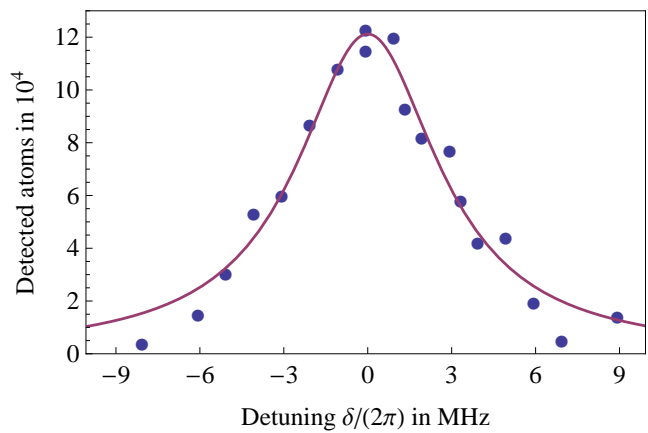}
\caption{(Color online) In-situ absorption imaging of ultra-cold atoms in an optical dipole trap. The laser was detuned by $\delta$ from the atomic resonance using the offset lock described in Section\,\ref{sll}, varying the detected atom number (circles). A Lorentzian of width $\Gamma_{\rm Fit} = 2\pi\times(6.1 \pm 0.4)\,{\rm MHz}$ is fitted to the data (solid line).}
\label{linewidth}
\end{figure}
The results are presented in Fig.\,\ref{linewidth}. A least-squares fit according to eq.\,(\ref{natoms}) results in a linewidth of $\Gamma_{\rm fit} = 2\pi\times(6.1 \pm 0.4)\,{\rm MHz}$, a value compatible with the natural linewidth of $2\pi\times(5.872 \pm 0.002)\,{\rm MHz}$ of \citep{McAlexander1996}. Within our experimental resolution we infer that the laser linewidth is much smaller than the natural linewidth of the atomic transition. Assuming a Lorentzian lineshape for the laser, the linewidth can be given as $200^{+400}_{-200}\,{\rm kHz}$, compatible with zero. %An upper bound for the laser linewidth can be given by the fit error of $400\,{\rm kHz}$.

\subsection{Long-term stability}

\begin{figure}
\includegraphics[width=1\columnwidth]{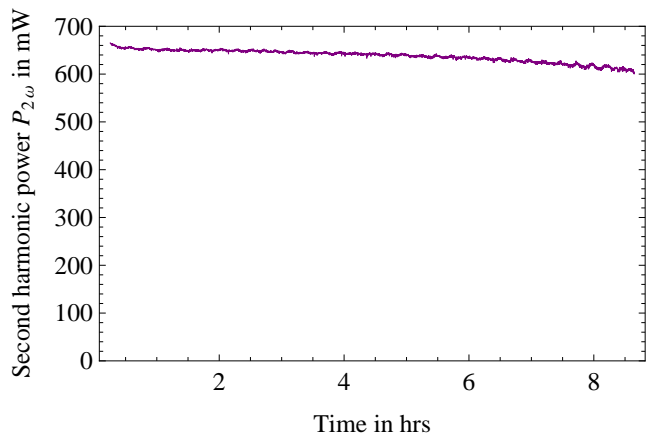}
\caption{(Color online) Long-term stability of the SH output power, experiencing a drop of 7\% over the measurement time. The system remained offset-locked to the lithium resonance.}
\label{f:longterm}
\end{figure}

Fig.\,\ref{f:longterm} shows a long-term stability plot of the laser system under laboratory conditions. The system remained locked during the measurement time of 8.5\,hours. The SH output power drops by 7\% and shows small modulations of a period of $\simeq 15\,\rm min$. This is attributable to slight angular tilts when the cavity's slow PZT (M$_4$/M$_3'$) is driven. This effect, changing the alignments, is confirmed by monitoring the laser output power, which drops by 5\% in the same time interval and displays the same modulations.

\section{Intra-cavity doubling}
\label{icd}

We also implemented the more direct approach of intra-cavity
doubling a 1342-nm laser. This concept simplifies the
optical design since only one cavity is needed. It was achieved by
using a setup similar to the one presented in Fig.\,\ref{laser}.
All cavity mirrors are highly reflective at $1342\,\rm nm$ and
mirror M$_3$ is also transmitting at $671\,\rm nm$. A nonlinear
crystal is put in the waist between mirrors M$_3$ and M$_4$.

For the Faraday rotator, we have tried various arrangements,
using either Gadolinium Gallium Garnet (GGG) or TGG as the Faraday material and
we have used either a rotatory power plate (made either of TeO$_2$ or of crystalline quartz) or a half-wave plate
to compensate the Faraday rotation. Although theory \citep{Biraben1979} favors the use of a rotatory power with respect
to a half-wave plate, we have found that the wave plate was more convenient, with a slightly larger output power.

\subsection{Infrared power}

\begin{figure}
\includegraphics[width=1\columnwidth]{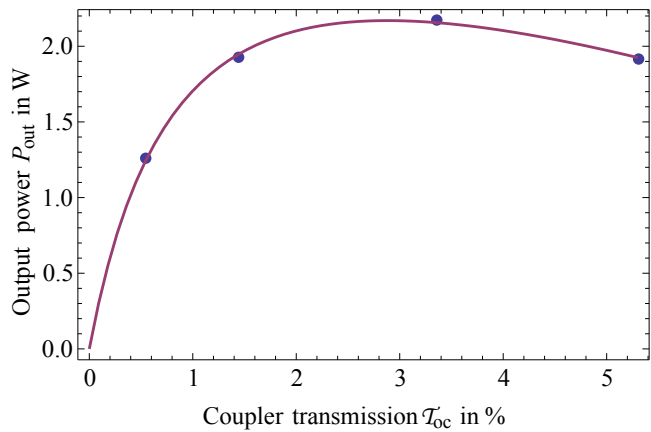}
\caption{ Output power $P_{\rm out}$ of the laser emitting at $1342\,\rm nm$ as a function
of the mirror transmission $\mathcal{T}_{\rm oc}$. The data points are experimental while the curve is the best fit using eq.\,(\ref{eqrigrod}).} \label{rigrod}
\end{figure}

In a setup involving intra-cavity frequency doubling, it is
essential to have very low parasitic losses $\mathcal{L}_{\rm
par}$ \citep{Smith1970}. We start by evaluating these losses by
measuring the emitted infrared laser power as a function of the
output mirror transmission $\mathcal{T}_{\rm oc}$ for a fixed
absorbed pump power $P_{\rm abs} =13\,\rm W$ for which thermal
effects in the Nd:YVO$_4$ crystal remain small. The data is
presented in Fig.\,\ref{rigrod}. Accurate values of the
transmission coefficient of the various output mirrors have been
obtained  with an absolute uncertainty near $0.03$\% by measuring
successively the direct and transmitted power of a laser beam in
an auxiliary experiment. According to
\citep{Chen1997,Rigrod1965,Laporta1991,Chen1996,Chen1999}, the
output power $P_{\rm out}$ is given by

\begin{eqnarray} \label{eqrigrod}
P_{\rm out} = P_{\rm sat} \mathcal{T}_{\rm oc}
 \left[\frac{G_0}{\mathcal{T}_{\rm oc}+\mathcal{L}_{\rm par}} -1\right]\,\rm ,
\end{eqnarray}
\noindent where $P_{\rm sat}$ is the gain medium saturation power
and $G_0$ the laser gain. We performed a nonlinear curve fit yielding $\mathcal{L}_{\rm par} = (0.0101 \pm  0.0006)$, $P_{\rm sat} = (26.3 \pm  2.0) \,\rm W$ and $G_0 = (0.150 \pm  0.006)$.
%\begin{eqnarray} \label{results}
%\mathcal{L}_{\rm par} &= (0.0101 \pm  0.0006)& \rm , \nonumber \\
%P_{\rm sat} &= (26.3 \pm  2.0)& \,\rm W\rm , \nonumber \\
%G_0 &= (0.150 \pm  0.006)& \rm . \nonumber
%\end{eqnarray}
%
%% this only works with amsmath
%\begin{alignat}{5} \label{results}
%\mathcal{L}_{\rm par} &={}& (0.0101& \pm {} & 0.0006&) \rm , \nonumber \\
%G_0                   &={}& (0.150& \pm {} & 0.006&) \rm , \nonumber \\
%P_{\rm sat}           &={}& (26.3& \pm {} & 2.0&) \,\rm W. \nonumber 
%\end{alignat}
%
%\begin{alignat}{5}
%    10a& ={}&  3x&& 3y& +{}& 18z&& 2w&\\
%     6a& ={}& 17x&&   & +{}&  5z&& 19w&
%\end{alignat}
The measured losses $\mathcal{L}_{\rm par}$ of
$\sim1\%$ are in accordance with expectations for a cavity made of four
mirrors (three high reflection mirrors plus the output mirror), three
AR-coated crystals and a Brewster plate. In Annex~\ref{Aaddmat} we relate the values measured for $P_{\rm sat}$ and $G_0$ to the parameters of the lasing crystal and the laser cavity.
We find good agreement with literature values.

\subsection{Doubling and frequency behavior}

Several Nd:YVO$_4$ lasers emitting at 671\,nm  have been built
based on intra-cavity frequency doubling using a LBO (lithium
triborate, LiB$_3$O$_5$) crystal
\citep{Agnesi2002,Agnesi2004,Lue2010,Ogilvy2003,Yao2004,Yao2005,Zhang2005,Zheng2002,Zheng2004}.
The largest achieved power was $5.5\,\rm W$ but none of these
lasers have run in SLM operation. We have tried frequency doubling
with both LBO and BIBO (bismuth triborate,
BiB$_3$O$_6$) crystals installed in the small waist of the laser cavity (see Fig. \ref{laser}). The BIBO crystal gave slightly more power
but with a substantially more astigmatic laser mode.
%\hl{(That's OK if M$^2$ is OK?! YOU ARE PROBABLY RIGHT BUT WE HAVE NOT MEASURED M$^2$ AND WE HAVE NOT TRIED TO CORRECT THE ASTIGMATISM OF THE OUTPUT BEAM)}.
%\hl{ ANSWER TO YOUR QUESTION :because of the thermal lens of the Nd:YVO$_4$ crystal, this waist is not exactly at mid-distance... Is it necessary to explain this point}.
Therefore we use a LBO crystal of $15\,\rm mm$ length and $3 \times 3 \,\rm mm^2$ cross-section. We apply type~I SH generation, with critical phase matching at $\theta = 86.1^{\circ}$ and $\phi =0^{\circ}$. % \hl{(should we define the angles? I HOPE NOT)}.
The crystal is AR coated with a {specified} residual reflection equal to $0.15$\% at 1342\,nm and  $0.5$\% at 671\,nm.

The non-linear optical coefficient of LBO is $d_{\rm eff} = 0.817 \,\rm
pm.V^{-1}$. Using the expressions given in ref.~\citep{Kontur2007} and the SNLO software
\citep{SNLO} to evaluate the crystal properties, we have calculated the expected optimum conversion
coefficient $\eta$ for this crystal.

We get  $\eta = 7.3 \times
10^{-5}\,\rm W^{-1}$ with an optimum waist in the crystal equal to
$w_0= 29\,\rm \mu m$. We use a slightly larger
 laser waist of $\simeq 45\,\rm \mu m$ for which theory predicts
$\eta = 4.9 \times 10^{-5}\,\rm W^{-1}$. We have measured $\eta$
by running the laser with a weakly IR-transmitting mirror M$_2$,
with a coupling transmission value $\mathcal{T}_{\rm oc} =
(0.55\pm0.03)\%$, and by measuring simultaneously the emitted
power at both 1342\,nm and 671\,nm. We have found $\eta = (4.7 \pm
0.5) \times10^{-5}\,\rm W^{-1}$, in excellent agreement with the
theoretical value.

Finally by replacing the IR-transmitting mirror M$_2$ by a highly
reflective one, we have extracted the SH output through mirror
M$_3$ which has a transmission near $95$\% at 671\,nm. When the
laser operates at the peak of its gain curve, corresponding to a
visible emission near $671.1\,\rm nm$, we get up to $1\,\rm W$ of SH
light.  At the lithium resonance wavelength {$670.8$}\,nm and
 with an intra-cavity  500-$\mu$m-thick
etalon made of fused silica and with a reflectivity of
$\mathcal{R} \simeq 30$\%, the current output power reaches $\sim
600\,\rm mW$. Progress towards frequency-stabilization as described in
Section\,\ref{sll} is ongoing. With this simpler optical system we
expect performances comparable to those obtained with external
cavity frequency doubling presented in Section\,\ref{dcav}.

\section{Conclusion}
\label{con} We have presented a frequency-stabilized laser source
to address the D-line transitions in atomic lithium. Up to
$670\,\mathrm{mW}$ single mode output power has been generated,
currently limited by intensity-dependent effects in the doubling
crystal. Tunability, narrow-band spectral quality and stable
long-term locked operation were demonstrated. This proves the
suitability of the system as a laser source for experiments
involving cooling and trapping of lithium species. We also
presented first results of a simpler alternative setup featuring
intra-cavity doubling. Higher output powers could be achieved by
optimizing the doubling cavity design for less intensity in the
ppKTP crystal by enlarging the waist or using less thermally
sensitive doubling nonlinear materials. Further increase of the
laser power at $1342\,\mathrm{nm}$ is feasible using a pump source
at $888\,\mathrm{nm}$ \citep{McDonagh2006}, reducing the quantum
defect and thus the detrimental thermal effects.

\begin{acknowledgements}
We would like to thank Y.~Louyer, P.~Juncar and F.~Balembois for their help in starting this development, J.L.~Doualan, R.~Moncorg\'e and P.~Camy for a new measurement of the stimulated cross section of Nd:YVO$_4$, E.~Mimoun for providing his design of the lock circuitry, N.~Navon for his help in conducting the linewidth measurements, F.~Lenhardt, P.~Zimmermann and J.J.~Zondy for fruitful discussions and R.~Stein and A.~Grier for careful reading of the manuscript. The LKB team acknowledges financial support from ERC (FerLoDim project), EuroQUAM (FerMix), IFRAF, IUF, ESF, ANR, SCALA and DARPA (OLE project) and the LCAR team acknowledges financial support from CNRS INP, BNM (grant 04-3-004), ANR (grant ANR-05-BLAN-0094) and R\'egion Midi-Pyr\'en\'ees.
\end{acknowledgements}

\appendix
\section{Additional materials}\label{Aaddmat}

The theoretical values of $P_{\rm sat}$ and $G_0$ are given by

\begin{eqnarray}
G_0 &=& \eta_{\rm Q} \eta_0 \frac{\lambda_{\rm p}}{\lambda_{\rm l}} \times \frac{P_{\rm abs}}{P_{\rm sat}}\rm , \\
P_{\rm sat} &=& \eta_0 I_{\rm sat} \frac{V_{\rm eff}}{L_{\rm
med}}\rm ,
\end{eqnarray}
\noindent where $\eta_0$ is the overlap efficiency of the pump and
laser cavity mode in the gain medium, $\eta_{\rm Q}$ the quantum
efficiency of emission, $V_{\rm eff}$ the gain medium effective
volume and $L_{\rm med}$ its length, $\lambda_{\rm p}$ and
$\lambda_{\rm l}$ the pump and laser wavelength respectively and
$P_{\rm p}$ the pump laser power. In
\citep{Chen1997,Rigrod1965,Laporta1991,Chen1996,Chen1999}\nocite{Chen1997,Rigrod1965,Laporta1991,Chen1996} the
length $L_{\rm med}$ is multiplied by $2$ because the calculation
concerns standing-wave cavities and this factor is suppressed in
the case of a ring cavity. Ref.~\citep{Chen1996} gives the general
expressions of $V_{\rm eff}/L_{\rm med}$ and of $\eta_0$ as a
function of the laser mode waist $w_{\rm l}$ and pump mode waist
$w_{\rm p}$.

In our experiment, the pump mode is obtained by expanding the mode
emitted by a 200-$\mu$m diameter optical fiber of
\mbox{$\mathit{NA}=0.22$} by a factor of $5$. After expansion, the
pump mode divergence is small and it is a reasonably good
approximation to assume that the pump mode waist is constant
over the crystal volume. We get

\begin{eqnarray}
\eta_0 &=& \frac{w_{\rm l}^2\left(w_{\rm l}^2+ 2 w_{\rm p}^2\right)}{\left(w_{\rm l}^2+ w_{\rm p}^2\right)^2} = 0.63\rm , \\
%\eta_0 &=& \frac{w_{\rm l}^2\left(w_{\rm l}^2+ 2 w_{\rm p}^2\right)}{\left(w_{\rm l}^2+ 2 w_{\rm p}^2\right)^2} = \hl{\frac{w_{\rm l}^2}{w_{\rm l}^2+ 2 w_{\rm p}^2} (=0.24)} = 0.63\rm ,  \\
\frac{V_{\rm eff}}{L_{\rm med}} &=& \frac{\pi}{2}\left(w_{\rm
l}^2+ 2 w_{\rm p}^2\right) = 6.4 \times 10^{-3}\,\rm {cm}^2\,\rm ,
\label{equationVeff}
\end{eqnarray}

\noindent where we have used for the laser mode waist the value
$w_{\rm l}=400\,\rm \mu m$ calculated at the position of the
Nd:YVO$_4$ crystal.  {This calculation assumes that the thermal
focal length of this crystal is $100$\,mm  but the mode parameters
are not very sensitive to this focal length, because its position
is close to the the large waist of the laser cavity}. The pump
mode waist $w_{\rm p}= 500\,\rm \mu m$ is deduced from the fiber
diameter and the expansion ratio.

By combining $G_0$ and $P_{\rm sat}$, we get $\eta_0 = 0.50 \pm
0.06$, reasonably close to our theoretical value. To get the value
of $V_{\rm eff}/L_{\rm med}$, we need to know the saturation
intensity $I_{\rm sat} = h \nu_{\rm l} /\sigma \tau_{e}$ where
$\sigma$ is the stimulated emission cross-section, $\tau_{e}$ the
excited state lifetime and $\nu_{\rm l} = c/ \lambda_{\rm l}$. The
maximum value of $\sigma$ for {stimulated} emission near
$\lambda_{\rm l}= 1342$ nm is $\sigma = 17\times 10^{-20}$ cm$^2$:
this value was measured with a spectral
resolution near $2.5\,\rm nm$~\citep{Fornasiero1998}  and the same value has been found in
an unpublished study \citep{Doualan2008}. The excited state
lifetime increases when the Neodymium ion concentration decreases
\citep{Okida2005,Tucker1977}. We have used the largest literature
value $\tau_{e} = 1.1\times 10^{-4}\,\rm s$ corresponding to a
$0.4$\%  Nd concentration. We thus get $ I_{\rm sat} = 7.9\pm
0.8\,\rm kW/cm^2$, with an estimated $10$\% error bar. The quantum
yield $\eta_{\rm Q}$ depends on the Neodymium ion concentration
\citep{Chen1999a} and practically $\eta_{\rm Q}=1$ for
0.2\,at.\%-doped crystals. By combining our measured values of
$G_0$ and $P_{\rm sat}$, we deduce $V_{\rm eff}/L_{\rm med} = (6.6
\pm 0.9) \times 10^{-3}\,\rm cm^2$ in good agreement with eq.\,(\ref{equationVeff}). This gives confidence in our determination of the $1\%$ loss of our cavity without second
 harmonic generation.

%\bibliographystyle{apalike}
%\bibliographystyle{unsrt}
%\bibliographystyle{apsrev}
%\bibliographystyle{nar}
\bibliographystyle{prstyNoEtAl}
\bibliography{Paper_red_laser}   % name your BibTeX data base

%% Non-BibTeX users please use
%\begin{thebibliography}{}
%%
%% and use \bibitem to create references. Consult the Instructions
%% for authors for reference list style.
%\end{thebibliography}

\end{document}